\documentclass[twocolumn,twocolappendix,trackchanges]{aastex63}
\usepackage{amsmath}
\newcommand{\code}[1]{\texttt{#1}}

\newcommand{\MESA}{\code{MESA}}

\usepackage{CJK}
\DeclareRobustCommand{\Eqref}[1]{Eq.~\ref{#1}}
\DeclareRobustCommand{\Figref}[1]{Fig.~\ref{#1}}

\DeclareRobustCommand{\Secref}[1]{Sec.~\ref{#1}}

\begin{document}

\title{The Stellar Merger Scenario for Black Holes in the
  Pair-instability Gap}
\author[0000-0002-6718-9472]{M.~Renzo}
\affiliation{Center for Computational Astrophysics, Flatiron Institute, New York, NY 10010, USA}
\affiliation{Department of Physics, Columbia University, New York, NY
  10027, USA}
\author[0000-0002-8171-8596]{M.~Cantiello}
\affiliation{Center for Computational Astrophysics, Flatiron Institute, New York, NY 10010, USA}
\affiliation{Department of Astrophysical Sciences, Princeton
  University, Princeton, NJ 08544, USA}
\author[0000-0002-4670-7509]{B.~D.~Metzger}
\affiliation{Department of Physics, Columbia University, New York, NY
  10027, USA}
\affiliation{Center for Computational Astrophysics, Flatiron Institute, New York, NY 10010, USA}
\author[0000-0002-2624-3399]{Y.-F.~Jiang (\begin{CJK*}{UTF8}{gbsn} 姜燕飞\end{CJK*})}
\affiliation{Center for Computational Astrophysics, Flatiron
  Institute, New York, NY 10010, USA}

\begin{abstract}
  The recent detection of GW190521 stimulated ideas on how to populate
  the predicted black hole {(BH)} pair-instability mass gap. One proposal is
  the dynamical merger of two stars below the pair instability regime
  forming a star with a small core and an over-sized
  envelope. {We outline the main challenges this scenario
    faces to form one BH in the gap. In particular, the core
    needs to avoid growing during the merger, and the merger product
    needs to retain enough mass, including} in the subsequent
  evolution, and at core-collapse. We explore this scenario with
  detailed stellar evolution calculations, starting with ad-hoc
  initial conditions enforcing no core growth during the merger. We
  find that these massive merger products are likely helium-rich and
  spend most of their remaining lifetime within regions {of
  instabilities in the Herzsprung-Russell diagram, such as luminous blue variable eruptions}.
  An energetic estimate of the amount of mass loss neglecting the
  back-reaction of the star suggests that the total amount of mass
  that can be removed at low metallicity is
  $\lesssim 1\,M_\odot$. This is small enough that at core-collapse
  our models are retaining sufficient mass to form black holes in the
  pair-instability gap similar to the recent ones detected by
  LIGO/Virgo. However, mass loss at the time of merger, {the
    resulting core structure}, and the mass loss at core collapse still need to be
  quantified for these models to confirm the viability of
  this scenario.
\end{abstract}
\vspace*{-10pt}

\section{Introduction}
\label{sec:intro}

The existence of a mass gap where pair-instability {(PI)} supernovae
\citep[SNe, e.g,][]{barkat:67} prevent the formation of black holes (BH) is a
robust prediction of stellar evolution theory
\cite[e.g.,][]{woosley:02, woosley:17, marchant:19, farmer:19,
  leung:19, renzo:20conv, renzo:20csm, marchant:20}. However,
the detection GW190521 \citep{GW190521a, GW190521b} is
challenging this prediction. For this merger event \emph{both} BH
masses\footnote{However, see also \cite{fishbach:20} for a
  population-informed re-analysis reconciling the masses
  with stellar evolution predictions.}, $8{5}^{+21}_{-14}\,M_\odot$
and $66^{+17}_{-18}\,M_\odot$, are in the PI mass
gap situated roughly between
$\sim$$45\,M_\odot$ and $\sim$$130\,M_\odot$
\citep[e.g.,][]{woosley:02, farmer:19, farmer:20}.

These BHs might not be the direct remnants of stars, but rather the
product of second-generation BH mergers in a (nuclear) cluster
\citep[e.g.,][]{Perna:2019,Rodriguez:2020,Mapelli:2020,Kremer:2020,Fragione:2020}
or AGN disk
\citep[e.g.,][]{McKernan:2012,McKernan:2014,Bartos:2017,Stone:2017,Fragione:2019},
{of accretion from gas clouds \citep[e.g.,][]{roupas:19, safarzadeh:20}}, or possibly
primordial BHs \citep{deluca:20}.

Many possible stellar explanations for the formation of BHs in
the gap have also been proposed. These include: reduction by
$\gtrsim 2\sigma$ of the
$^{12}\mathrm{C}(\alpha,\gamma)^{16}\mathrm{O}$ reaction rate
\citep{farmer:20, belczynski:20b}, modifications to the standard model
\citep{croon:20a, croon:20b, sakstein:20}, and population III stars
\citep{farell:20, kinugawa:20}.  BHs accreting in isolated binaries do not
contribute significantly to {populating} the pair-instability mass
gap \citep{vanson:20}.

We focus on one particular scenario proposed by
\citet{spera:19, dicarlo:19, dicarlo:20a, dicarlo:20b}: the formation of BHs in the PI mass gap via stellar mergers in a dynamical environment.
\Secref{sec:scenario} summarizes this
``stellar merger'' scenario and its challenges.
We then construct stellar models for the merger product in \Secref{sec:methods} and evolve them.
\Secref{sec:results} shows how evolutionary processes can lead to continuum-driven
mass-loss which is not considered in rapid population-synthesis
models underlying N-body calculations. We {estimate the amount
of mass loss}, and conclude that by itself it might not change the
scenario appreciably.

\section{The stellar merger scenario}
\label{sec:scenario}

\cite{dicarlo:20b} present a detailed example of this scenario (see
their Fig.~7), where the dynamically-driven merger happens at the end
of the main sequence of a
$\sim$$58\,M_\odot$ star. At this point the total mass of a
$\sim$$42\,M_\odot$ star is added to the envelope \emph{without}
modifying the core.
Thus, the merger product has a small core and an over-sized envelope,
and reaches core-collapse (CC) with a total mass of
$\sim$$99\,M_\odot$. At this point, \cite{dicarlo:19, dicarlo:20a, dicarlo:20b} assume a direct collapse to BH, without mass ejection but accounting for a reduction in gravitational mass due to neutrino losses.

In order to form a coalescing binary BH with masses in the PI mass gap, this scenario faces the following four
challenges.

\subsection{The merger challenge: mass loss{, core size,} and rotation}

The first challenge is retaining sufficient mass during the stellar
collision {and forming a post-merger structure without
  modifying the core}.

\citet{lombardi:02} performed SPH simulations of low-mass
stellar collisions and found their models lose 1--7\,\% of the total
mass during the mergers.  \citet{glebbeek:13} computed SPH simulations
of head-on collisions of massive stars. For their most massive objects
(40$\,M_\odot$+40$\,M_\odot$) they found a mass loss corresponding to
6--8\,\% of the total mass. However, these models neglect the effects
of radiation transport, which could have an important role for the
mass budget in the merger of very luminous stars. Including radiation
effects would likely increase the mass-loss rate during the merger,
because the radiation-pressure dominated envelope of very massive
stars are loosely bound and easily stripped.

A back-of-the-envelope estimate of the amount
of mass loss can be obtained considering the energy available to drive
mass loss is (a fraction of) the relative kinetic energy
of the two incoming stars
\begin{equation}
  \label{eq:E_kin}
  E_\mathrm{kin}\sim \frac{1}{2} \frac{M_1M_2}{M_1+M_2} v_\sigma^2 \lesssim 10^{46}\,\mathrm{erg} \ll E_\mathrm{bind} \ \ ,
\end{equation}
where we use the aforementioned masses and assume
$v_\sigma\lesssim 10\,\mathrm{km\ s^{-1}}$ as the velocity dispersion
of the cluster, and $E_\mathrm{bind}$ is the typical binding energy of
the stars. {Our pre-merger models (see \Secref{sec:methods}) only have
  $\lesssim 10^{-3}\,M_\odot$  with binding energy lower than $10^{46}$\,erg.} This
suggests that mergers resulting in small mass loss might be possible,
{however, using instead the escape velocity from the star in
  $E_\mathrm{kin}$ would result in a significantly higher mass loss.}  Our estimate neglects the stellar
reaction to the energy injection during the merger process.

{A second challenge is maintaining the core mass below the
  pulsational pair-instability (PPI) regime. The result of a stellar
  collision is often approximated using entropy sorting
  \citep[e.g.,][]{lombardi:02, gaburov:08}: this would
  effectively result in merging the cores of both stars and increasing the
  resulting core mass. Dynamical interaction might also pair stars in tight
  binaries merging later in a common envelope event. Merger
  simulations involving one evolved and one unevolved star are often invoked to explain
  the progenitor of SN1987A \citep[e.g.,][]{podsiadlowski:92,
    menon:17}, and in this case mixing into the core decreasing its
  mass has been proposed.}

Mergers {products} are also expected to be fast rotators \citep[e.g.,][]{demink:13}, although
\cite{schneider:19} found {internal redistribution of
  angular momentum preventing fast surface rotation}. If a large amount of
angular momentum is transported in the core, rotational mixing
\citep[e.g.,][]{maeder:00} might
increase the core mass pushing the star into the PPI regime.

{We do not investigate how realistic is the merger structure
  proposed by \cite{spera:19,dicarlo:20a, dicarlo:20b}. Following
  these studies,} we assume
  no mass is lost during the merger process and we do not consider
  the effects of rotation.

\subsection{The evolution challenge: winds and envelope instabilities}

After the merger, low metallicity is necessary to prevent large line-driven
wind mass loss \citep[e.g,][]{farell:20,kinugawa:20}, or these have to be artificially
suppressed \citep{belczynski:20}. The angular momentum
distribution might also lead to centrifugally-driven mass loss
\citep{langer:98, heger:00, zhao:20}.

Other modes of mass loss such as continuum-driven winds and/or luminous blue
variable (LBV) eruptions are typically not considered. However, merger
products are prime candidates to explain LBV stars
\citep[e.g.,][]{justham:14, aghakhanlootakanloo:17} because of their increased luminosity,  non-standard
internal structure, and possible He-enrichment. He opacity is thought
to have a key role in driving  eruptive mass loss \citep{jiang:18},
which could make this type of mass loss relatively metallicity-independent.

Our simple models presented in \Secref{sec:results} address mainly
this problem at metallicity $Z=2\times10^{-4}$ \citep[e.g.,][]{dicarlo:20a},
however this challenge is expected to become progressively harder
at higher $Z$, because of the increasing opacity in the stellar envelope.
\newpage
\subsection{The collapse challenge: mass loss at black hole formation}

At BH formation,
$\sim$$10^{53}\,\mathrm{erg}$ of neutrino emission is expected
to suddenly decrease the gravitational mass of the collapsing core. This in turns creates a shock propagating thorough the envelope that can unbind the outer-layers
\citep[e.g.,][]{nadezhin:80, lovegrove:13, fernandez:18}. While in the calculations of \cite{dicarlo:20b} the
BH mass accounts for the neutrino losses, the impact on the envelope mass loss was not investigated.

If the envelope is not  lost at core collapse, it still could retain enough angular momentum to allow for the formation of an accretion disk.
This could result in (ultra)-long gamma-ray bursts \citep[e.g.,][]{perna:18} and the delayed ejection of a significant fraction of the envelope. Even in the absence of net rotation, the fallback of large convective cells
in the envelope could also drive the formation of disks and ultimately produce large amounts of mass loss through jets \citep{quataert:19}.

\subsection{The gravitational-wave challenge: dynamical pairing}

If the previous challenges can be overcome, this scenario allows for
the formation of single BHs in the gap. Dynamical interactions are
then required to pair two of these BHs together in a tight orbit.
Because of their large masses, these BHs are
efficient at finding companions to merge with, with typical delay-time distribution of order tens
of Myrs. \cite{dicarlo:20a} finds a significant fraction of the merger
rate from cluster dynamics to involve one BH from the stellar merger
scenario (see also \citealt{spera:19,dicarlo:19, Kremer:2020,
  dicarlo:20b}). However, to explain the masses in GW190521, both
BHs should have formed via such evolutionary path.

\section{Constructing a merger model}
\label{sec:methods}

We use Modules for Experiments in Stellar Astrophysics
\citep[\MESA~revision 12778,][]{paxton:11, paxton:13, paxton:15,
  paxton:18, paxton:19} to construct a post-merger structure. We do not
compute the dynamical phase of the merger, but rather construct an
ad-hoc post-merger structure starting from the pre-merger
stars. Details of the numerical implementation, together with the
microphysics inputs are given in
Appendix~\ref{sec:software}.

\subsection{Initial chemical composition}

\begin{figure}[tbp]
  \includegraphics[width=0.5\textwidth]{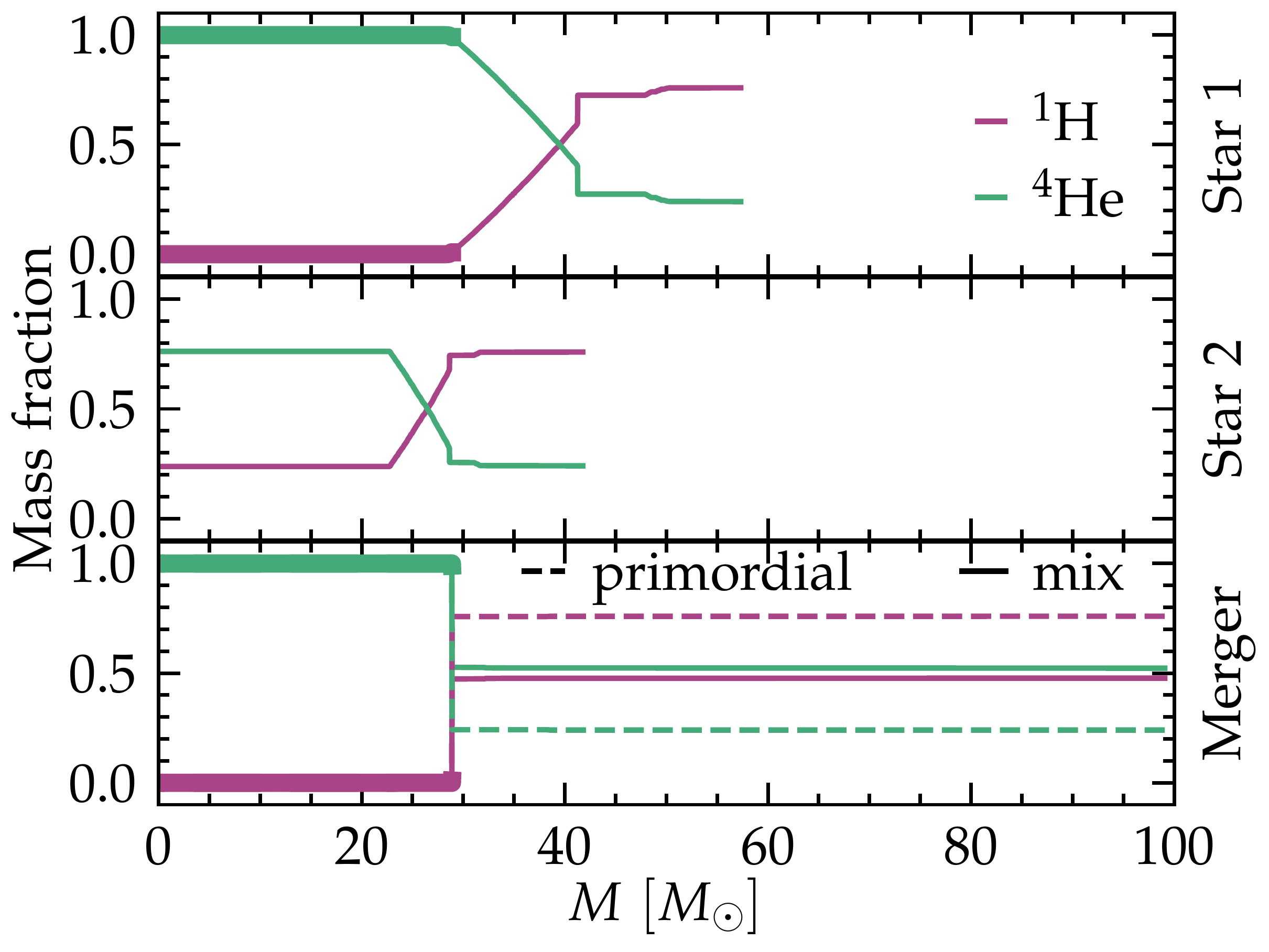}
  \caption{H and He profiles of the two pre-merger stars (top and
    middle panel) and of the merger products (bottom
    panel). {In the bottom panel, solid (dashed) lines
      indicate the envelope composition for the ``mix''
      (``primordial'') model. Both models have by construction the
      same core structure of the most massive star pre-merger (thicker
      lines). The least massive star 2 is too young to have a well
      defined He core.  }}
  \label{fig:composition}
\end{figure}

Following \cite{dicarlo:20b}, we assume the merger happens at the end
of the main sequence of a $M_1=58\,M_\odot$ star at
$Z=2\times10^{-4}$. Using \cite{brott:11} overshooting, its
main-sequence lifetime is $\tau_\mathrm{MS}=4.15\,\mathrm{Myr}$, and
the corresponding Helium (He) core mass is
$\sim$$29\,M_\odot$. This value is below the limit for any kind of PI pulse \citep{renzo:20csm}. We evolve up to
$\tau_\mathrm{MS}$ a
$M_2=42\,M_\odot$ star with the same setup. Very massive stars have comparable lifetimes, and by this time, the second star has a central He abundance
$X_c(^4\mathrm{He})=0.76$ extending out to mass coordinate
$\sim25\,M_\odot$ (cf.~middle panel of \Figref{fig:composition}). {The Kelvin-Helmholtz timescale of these stars is of order
  $10^4$\,years. After the merger, a relaxation phase of comparable duration is expected (although both the luminosity and radius are likely to be higher right after the merger, e.g., \citealt{schneider:19}).}

To construct the merger product, we first relax
\citep[e.g.,][]{morozova:15, vigna-gomez:19} the most massive star
model to $M_\mathrm{tot}=M_1+M_2-\Delta M_\mathrm{wind}$, where the
mass lost to winds $\Delta M_\mathrm{wind}$ is only
$\sim$$0.9\,M_\odot$ using \cite{vink:01} algorithm.
{The mass relaxation procedure does not account for
  the release of gravitational or
  internal energy from the newly accreted mass}.
Then, we relax the chemical composition of the merger.

\begin{figure*}[htbp]
  \includegraphics[width=0.5\textwidth]{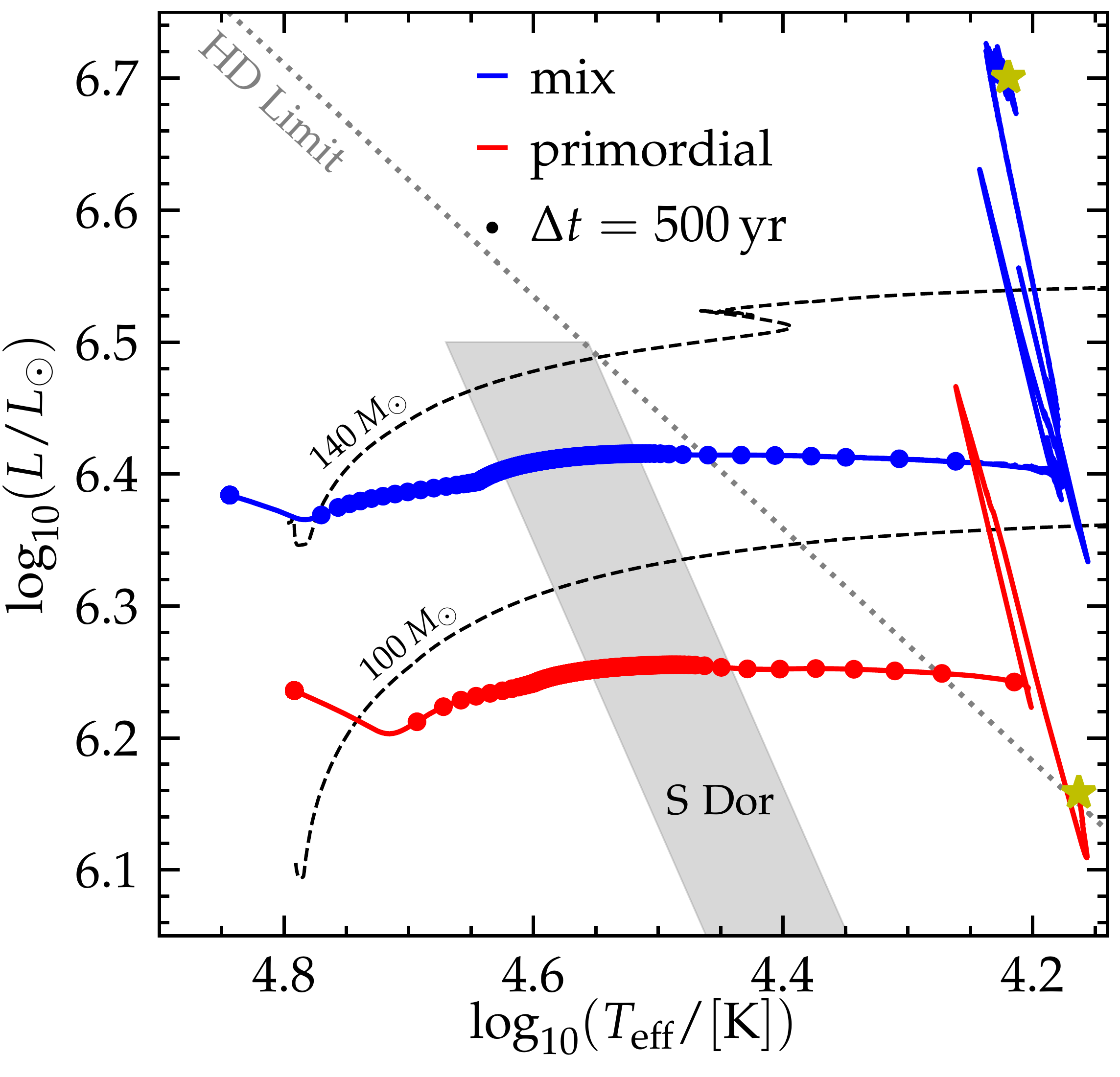}
  \includegraphics[width=0.5\textwidth]{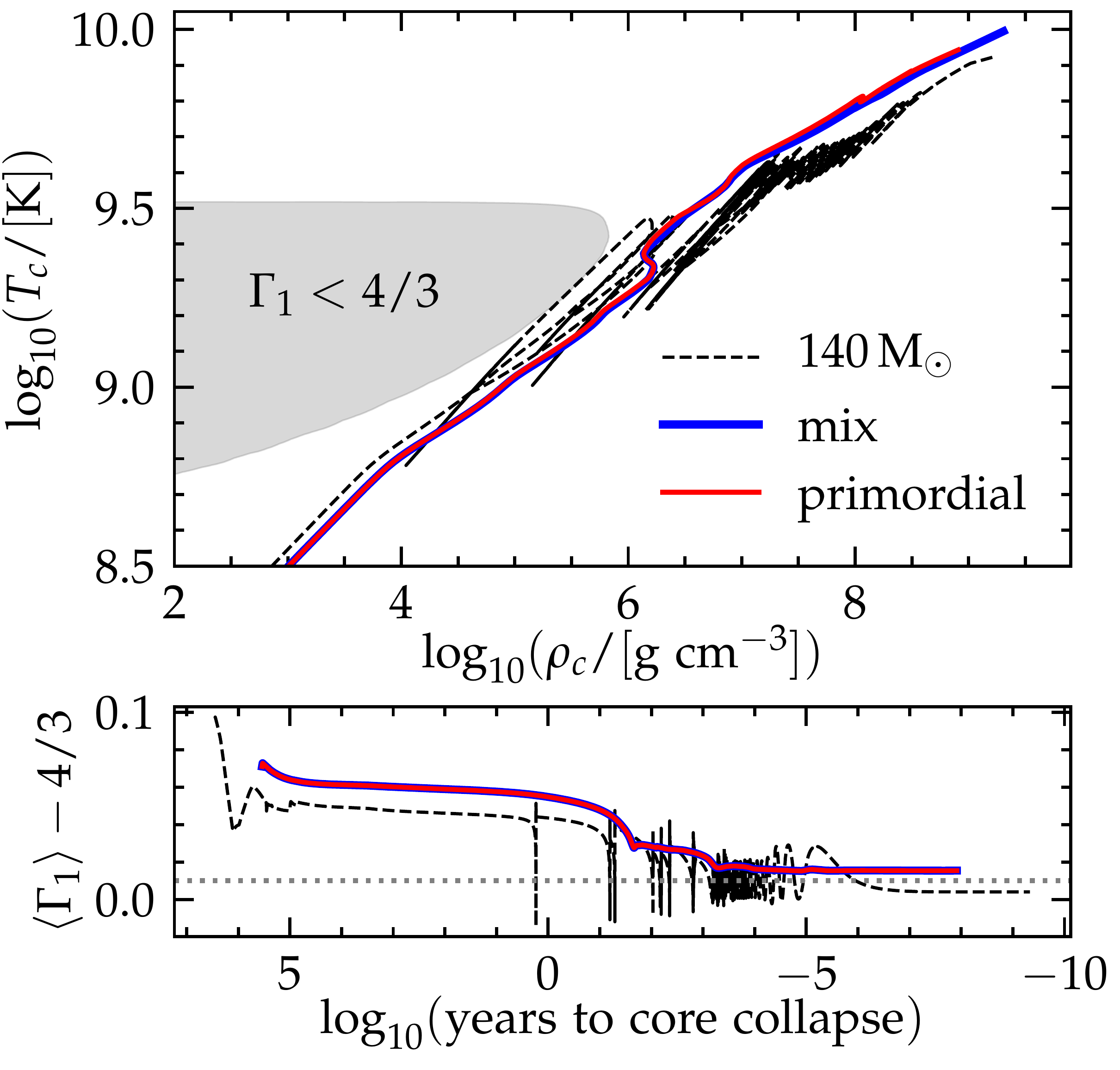}
  \caption{Left panel: HR diagram of the post-merger evolution. Each
    dot is separated by 500\,years. The dashed black lines shows for
    comparison a H-rich $140\,M_\odot$ star from \citealt{renzo:20csm}
    (smaller overshooting, encounters the PPI later), and a
    $100\,M_\odot$ model (same overshooting, expected to encounter the
    PPI).  The yellow stars mark the onset of core collapse. Top right
    panel: evolution of the central temperature and density.  Bottom
    right panel: time evolution of the pressure-weighted average
    adiabatic index, the star becomes pair-unstable when it drops below the
    dotted horizontal line \citep{renzo:20csm}.}
  \label{fig:evol}
\end{figure*}

The top two panels in \Figref{fig:composition} show the pre-merger
composition of the two stars, and the bottom panel shows two different
merger products.  For both, we enforce the hypothesis of the
``stellar merger scenario'' maintaining the same composition and mass of
the core {of the most massive star} (thick lines in
\Figref{fig:composition}).

The fact the second star has already synthesized a large amount of
$^{4}\mathrm{He}$ forces to make choices for the envelope composition.
Usually, mergers are built assuming that the lowest entropy layers
sink to the bottom. However, this would result in a larger He core
mass, entering the PPI regime, and violating
the hypothesis of the scenario.  Instead, in model ``mix'' (solid
lines in the bottom panel of \Figref{fig:composition}) we
fully mix the envelope of the most massive star with the entire
second star at merger time. This represents the most favorable scenario preventing
growth of the He core, leading to the formation of a He-rich envelope with
$X(^{4}\mathrm{He})\simeq0.52$. The total mass in each element is
conserved to better than $2\%$.

As the opposite limiting case, in model ``primordial'' (dashed lines
in the bottom panel \Figref{fig:composition}) we ignore the
composition of the second star, and increase the envelope mass with
the initial composition, that is with a He abundance of
$X(^{4}\mathrm{He})\simeq0.24$. This corresponds to a merger between
an evolved primary and a newly formed secondary star with its
primordial chemical composition.

\section{Post-merger evolution}
\label{sec:results}

We evolve our merger models until the onset of CC.
The left panel of \Figref{fig:evol} shows their post-merger Hertzsprung-Russell (HR)
diagram. The evolution proceeds from left towards cooler
temperatures. The more He-rich ``mix" model has a higher
luminosity ($L$). This can be understood considering that $L\propto\mu^4$
where $\mu$ is the mean molecular weight for an ideal gas with
constant opacity \citep[e.g.,][]{kippenhahn:13}.  Most of the He and
carbon core burning happens within the hot S Doradus instability strip
(S Dor, gray band in \Figref{fig:evol}), where the star spends
$1.9\times10^5\,\mathrm{years}$ for model ``mix'' and
$8.1\times10^4\,\mathrm{years}$ for model ``primordial".

Afterwards, both evolve into the observationally forbidden
region beyond the Humphrey-Davidson limit (HD, dotted gray line in \Figref{fig:evol},
\citealt{humphreys:94}). Model ``mix"" spends there its last
$\sim3800\,\mathrm{years}$, while owing to its lower luminosity,
the model ``primordial" only spends $\sim$$600\,\mathrm{years}$
beyond the HD limit.

While both the S Dor strip and the HD limit
have been observationally determined at  $Z \approx Z_\odot$,
\citet{Davies:2018} recently showed that the empirical HD limit  is likely
metallicity-independent. This might support the theoretical results of \citet{jiang:18}, who found that He-opacity
is the likely driver of outbursts in luminous stars close to the HD limit and the S Dor strip.
Overall, LBVs are known at the metallicity of the Small Magellanic Cloud
\citep[e.g,][]{szeifert:93}, and observation of narrow-lined SNe (in
particular their isolation relative to other explosions) might be
compatible with LBV eruptions happening in more metal-poor parts of
galaxies.

The location of our merger models on the HR diagram suggests they could be affected by
envelope instabilities and severe mass loss. The noisiness of the curves is caused by the numerical
instabilities, possibly related to physical
instabilities in the envelopes (see \Secref{sec:estimate_LBV}).

The top right panel of \Figref{fig:evol} shows the evolution of the
central temperature and density. By construction, both merger models
avoid the instability region (gray area) and proceed to CC avoiding
pulses. Conversely, a $140\,M_\odot$ single star  model, hits repeatedly the PI
(although off-center, \citealt{renzo:20csm}) resulting in large mass
loss. The bottom right panel shows the averaged adiabatic index $\Gamma_1$
which stays above the dotted line indicating instability \citep[e.g.,][]{renzo:20csm}. Conversely,
the averaged adiabatic index of the $140\,M_\odot$ model drops significantly
below 4/3 repeatedly during PI pulses.

\subsection{Estimates for the continuum-driven mass loss}
\label{sec:estimate_LBV}

\begin{figure}[tbp]
  \includegraphics[width=0.5\textwidth]{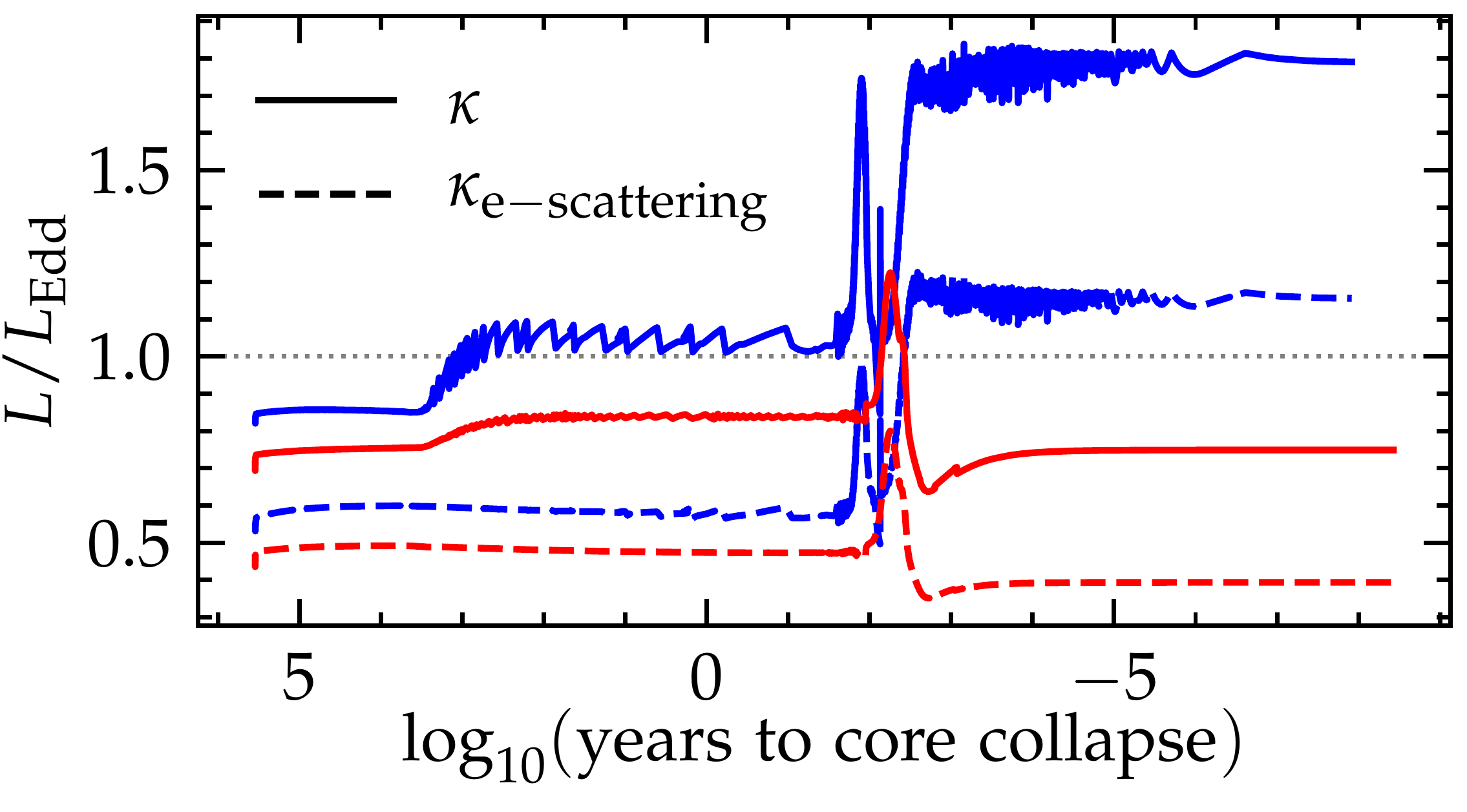}
  \caption{Eddington ratio post-merger as a function of the time left
    to CC. Solid lines use only the electron-scattering opacity for
    $L_\mathrm{Edd}$, while the dashed lines use the total opacity.  The
    blue (red) lines correspond to the ``mix'' (``primordial'') merger
    models. In both cases, the post-merger model exceeds Eddington
    ratio of 1 (dashed horizontal line), indicating that eruptive
    and/or continuum driven mass loss could occur.}
  \label{fig:Edd_ratio}
\end{figure}

\Figref{fig:Edd_ratio} shows the temporal evolution of the ratio of
the luminosity $L$ to the
 Eddington luminosity
\begin{equation}
  \label{eq:L_Edd}
  L_\mathrm{Edd}  =\frac{4\pi G M c}{\kappa} \ \ ,
\end{equation}
where $G$ is the gravitational constant, $M$ the total mass, and $c$ the
speed of light. Dashed lines only consider electron-scattering
opacity $\kappa\simeq0.2 (1+X(^1\mathrm{H}))\,\mathrm{cm^{2}\ g^{-1}}$, while solid lines correspond to using the
  stellar surface\footnote{The surface opacity is the average of the Rosseland mean opacity from optical depth $\tau = 2/3$ down to
$\tau=100$.} opacity in the calculation of the Eddington
luminosity.

Both our merger models evolve with high Eddington ratios, and the more
luminous and He-rich ``mix'' model reaches $\sim${1} about a
thousand years before CC.  Again, this suggests that
radiatively-driven eruptive mass loss might {occur even at}
low metallicity \citep[e.g.,][]{smith:14}. Increasing $Z$ has a large
effect over the opacity $\kappa$, decreasing $L_\mathrm{Edd}$ and thus
increasing the Eddington ratio. Not surprisingly, preliminary
calculations at $Z=0.02$ with our simple setup proved to be
numerically unstable.

\begin{figure}[htbp]
  \includegraphics[width=0.5\textwidth]{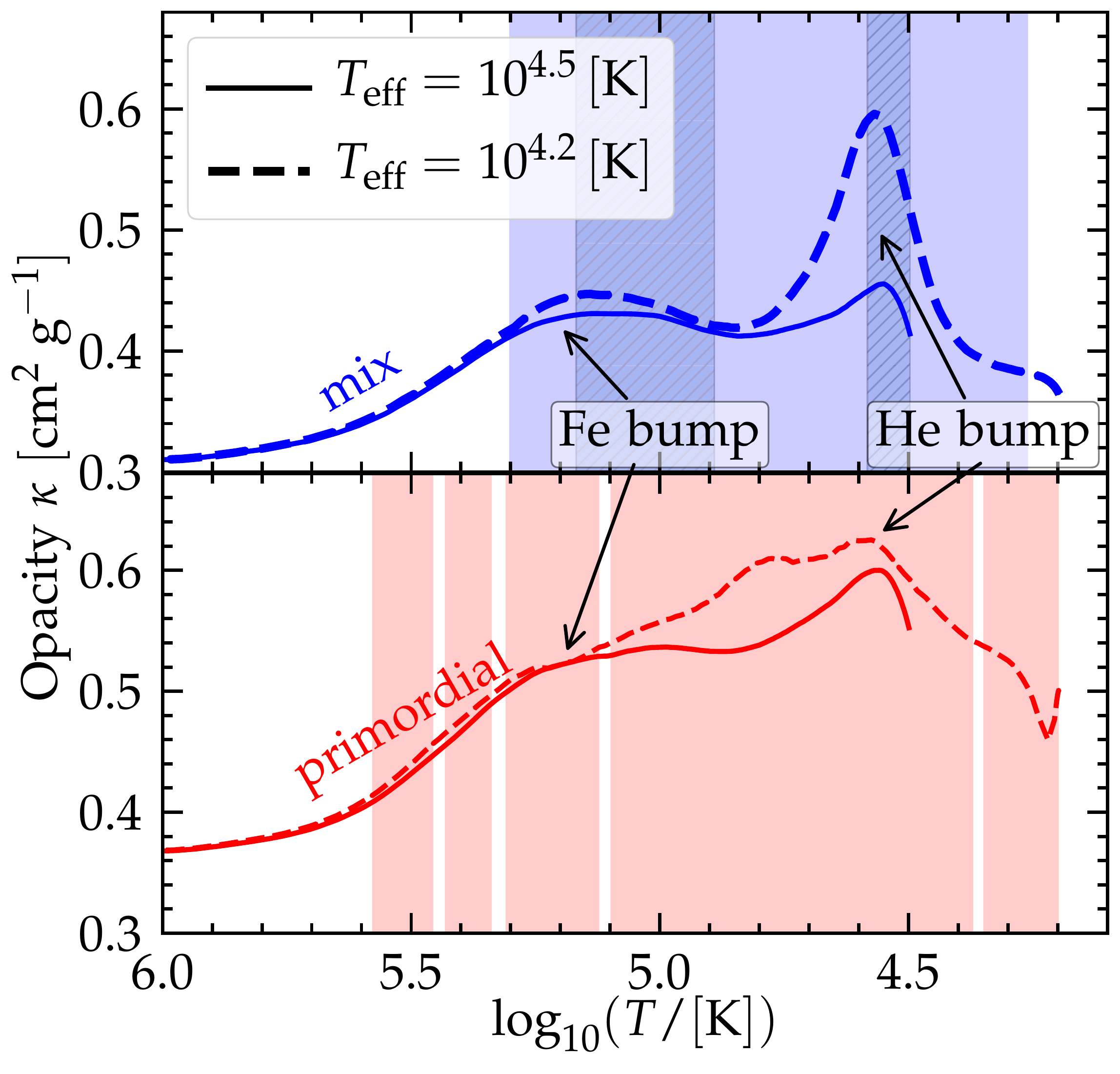}
  \caption{Outer structure of the opacity for two different values of the
    effective temperature.
    {The top (bottom) panel shows the ``mix'' ( ``primordial'') model. The hatched regions indicate
    convection at $\log_{10}(T_\mathrm{eff}/\mathrm{[K]})=4.5$, with
    two separate regions in the ``mix'' model (and none in the
    ``primordial'' model). Colored regions mark
    convective regions at $\log_{10}(T_\mathrm{eff}/\mathrm{[K]})=4.2$.}}
  \label{fig:opacity}
\end{figure}

\Figref{fig:opacity} shows the internal structure and opacity
profile at two selected times for our models. Solid lines correspond
to when the models reach an effective temperature $\log_{10}(T_\mathrm{eff}/\mathrm{[K]})=4.5$,
roughly in the S~Dor instability strip, while dashed lines show
models at $\log_{10}(T_\mathrm{eff}/\mathrm{[K]})=4.2$, beyond the HD
limit.

While our models have $Z=2\times10^{-4}\simeq Z_\odot/100$, two opacity bumps are still
evident at both times, one caused by partial recombination of iron
(Fe) roughly at $\log_{10}(T/\mathrm{[K]})\simeq5.3$, and one due to partial He
recombination at $\log_{10}(T/\mathrm{[K]})\simeq4.6$. The presence of
these opacity bumps drives inefficient convection layers (shading {and hatching} in
\Figref{fig:opacity}). The interplay between density
inhomogeneities due to convection and the
close-to-super-Eddington luminosity was found to be a key driver of
LBV eruptions (at least at $Z=Z_\odot$, \citealt{jiang:15, jiang:18}).

{Because of the as-yet insufficient theoretical understanding,
  o}ur models do not include eruptive LBV-like
mass loss or continuum-driven winds
in addition to line-driven winds. Nevertheless, following \citet{paxton:11,cantiello:20}
we can estimate the extra mass loss of models exceeding the Eddington luminosity as

\begin{equation}
  \label{eq:edd_wind}
  \dot{M}_\mathrm{Edd} \simeq -\frac{L-L_\mathrm{Edd}}{v_\mathrm{esc}^2} \ \ ,
\end{equation}
where $L$ and $L_\mathrm{Edd}$ are the luminosity and Eddington
luminosity, and $v_\mathrm{esc}$ is the surface escape velocity. We
only use this to estimate in post-process the amount of mass the star
would have lost, and we neglect the structural reaction this may cause
\citep{renzo:17}.

\begin{figure}[htbp]
  \includegraphics[width=0.5\textwidth]{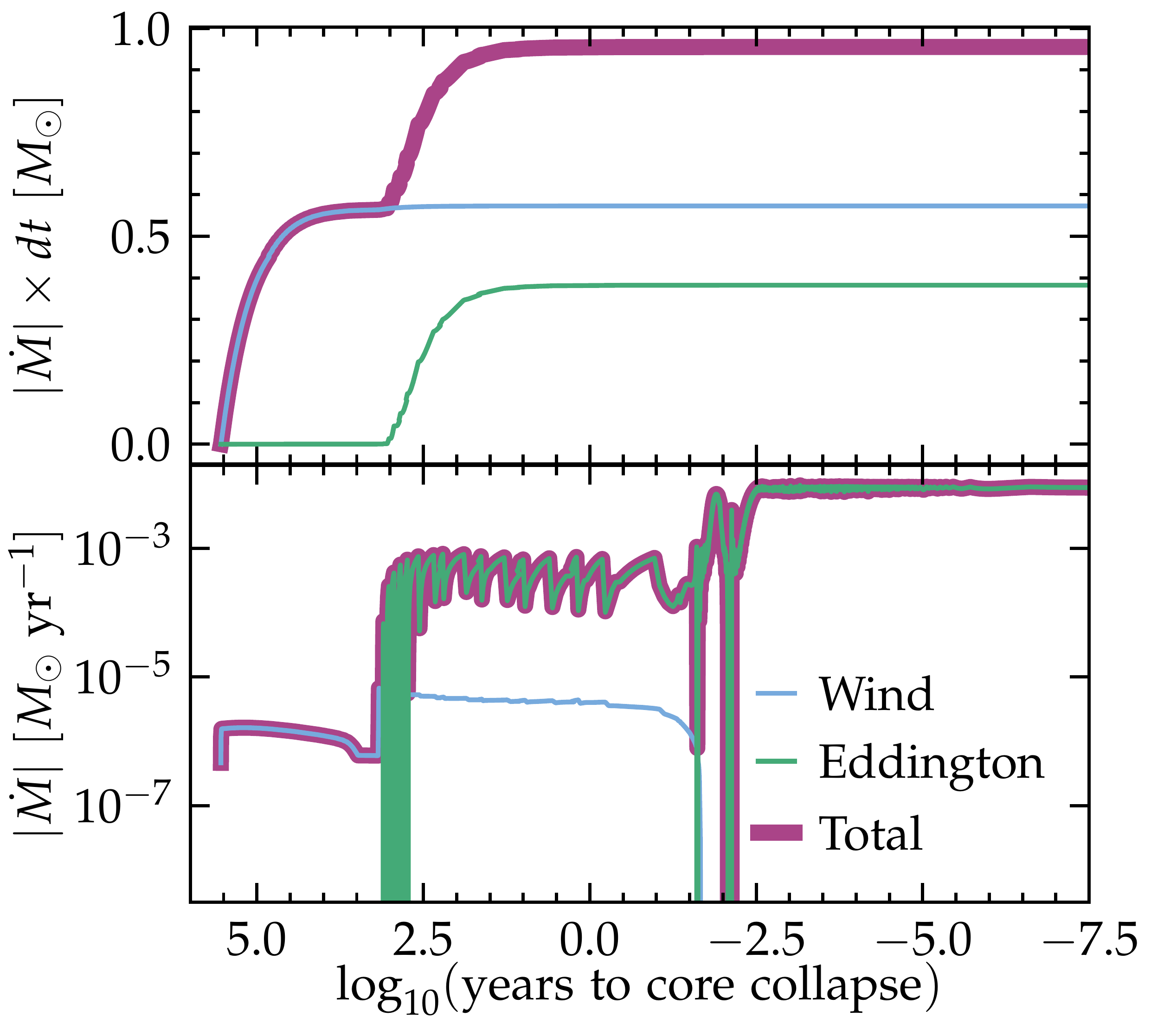}
  \caption{Absolute value of the mass loss rate (bottom) and cumulative mass
    loss (top) as a function of time until CC for our ``mix'' model.}
  \label{fig:mdot_estimate}
\end{figure}

\Figref{fig:mdot_estimate} shows the mass-loss {rate} history
({bottom} panel) and cumulative mass los{t}
({top} panel) for our ``mix'' model. The wind mass-loss rate
\citep[in blue,][]{vink:01} removes mass earlier on but becomes
subdominant a few hundred years before CC. Then, the Eddington-driven
mass loss (green, \Eqref{eq:edd_wind}) becomes dominant. The total
mass loss (thick purple) is the sum of the two and is only about
$\sim$$1\,M_\odot$, {corresponding to a total final mass of $\sim$$98\,M_\odot$}. At higher
$Z$ more mass loss would be expected.  The lower luminosity and Eddington ratio of the ``primordial'' model result in a smaller mass loss estimate than for the ``mix'' model.

\newpage
\section{Discussion \& conclusions}

We tested a stellar merger scenario for the production of BHs in the pair-instability mass gap.
To avoid the pair-instability regime, this scenario assumes that the core mass of
the primary star is unaffected by the merger,
with the mass of the secondary fully mixed into the envelope
\citep{spera:19, dicarlo:19, dicarlo:20a, dicarlo:20b}. We do not explore how realistic this assumption is, which needs to
  be addressed using hydrodynamic calculations \citep[e.g.,][]{lombardi:02, glebbeek:13, schneider:19}. Standard
entropy-sorting post-merger structures would result in adding the He
cores together, which would violate the hypothesis allowing these
mergers to avoid the PPI regime.

Assuming the core does not grow, the envelope of the merger
necessarily becomes He-enriched, with the extent of the enrichment
depending on when the merger occurs during the main sequence of the
secondary star. It could be prevented by allowing the least massive
star in the merger to be younger (and less evolved) than the most
massive one (e.g., our ``primordial'' model), so that less He is
present in its core. Given the small lifetime differences between very
massive stars, this would require not only a non-starburst star
formation history, but also fine-tuned timing.

We use detailed stellar evolution calculations to evolve two
merger products that assume either an evolved or an unevolved secondary.
Because of the He-enrichment, the merger product can be
significantly more luminous than a star of similar mass
(cf.~\Figref{fig:evol}). It is already possible most stars with
$M\gtrsim100\,M_\odot$ will experience large outbursts of mass loss
\citep[e.g.,][]{conti:75}, {and} the He-richness might exacerbate this
\citep{jiang:18}. This suggests LBV-like outbursts or continuum driven
mass-loss can occur even at
metallicity as low as $Z=2\times10^{-4}\simeq Z_\odot/100$.

We estimated the amount of mass that can be lost by these
stellar merger products due to their proximity to the Eddington limit,
and found that they can shed up to $\approx 1M_\odot$ during the last
few hundred years prior to core collapse (\Secref{sec:estimate_LBV}).
This circumstellar material could leave visible imprints in the light
curves and spectra of a terminal transient \citep[e.g.,][]{arcavi:17SN,
  vigna-gomez:19}, if the final BH formation ejects (a small amount
of) mass \citep[e.g.,][]{gilkis:14, quataert:19}.

However, the amount of material lost during the evolution is not large
enough to affect the scenario for BH formation in the pair-instability
mass gap. We note that stronger mass loss is expected in more
metal-rich environment, so that this scenario can only occur below a
metallicity threshold.  Ultimately, multidimensional radiation
hydro-dynamical simulations and a better theoretical understanding of
LBV eruptions is needed to precisely quantify the pre-collapse mass of
these luminous, He-rich merger remnants.

Finally an estimate of neutrino-driven mass loss \citep{nadezhin:80, lovegrove:13} is required to establish the actual size of the BH formed at core collapse.
The oversized envelopes and He-enrichment keep our merger models
relatively blue at the onset of core-collapse
($\log_{10}(T_\mathrm{eff}/\mathrm{[K]})\gtrsim 4.1$), that is in the intermediate
regime where the amount of mass loss at BH formation is unclear \citep[e.g.,][]{fernandez:18}.
Further studies of the hydrodynamics at merger and at BH formation are
needed to assess whether the ``stellar merger scenario'' can
contribute a significant populations of BHs inside the PISN mass
gap.

In the meantime, {population synthesis} simulations could bracket the
range of possibility by considering varying degrees of envelope mass
loss before and at core-collapse.

\software{
  \texttt{mesaPlot} \citep{mesaplot},
  \texttt{mesaSDK} \citep{mesasdk},
  \texttt{ipython/jupyter} \citep{ipython},
  \texttt{matplotlib} \citep{matplotlib},
  \texttt{NumPy} \citep{numpy},
  \MESA \citep{paxton:11,paxton:13,paxton:15,paxton:18,paxton:19}
}

\section*{Acknowledgements}

We are thankful to U.~N.~di~Carlo, R.~Fernandez, Y.~G\"otberg,
Y.~Levin, and N.~Smith for helpful exchanges, and to the referee for
the prompt and constructive feedback.

\appendix

\section{\texttt{MESA} setup}
\label{sec:software}

We use \code{MESA} version 12778 to compute our models.  The
\code{MESA} {equation of state} (EOS) is a blend of the OPAL \citet{Rogers2002}, SCVH
\citet{Saumon1995}, PTEH \citet{Pols1995}, HELM \citet{Timmes2000},
and PC \citet{Potekhin2010} EOSes.

Radiative opacities are primarily from OPAL \citep{Iglesias1993,
  Iglesias1996}, with low-temperature data from \citet{Ferguson2005}
and the high-temperature, Compton-scattering dominated regime by
\citet{Buchler1976}. Electron conduction opacities are from
\citet{Cassisi2007}.

Nuclear reaction rates are a combination of rates from NACRE
\citep{Angulo1999}, JINA REACLIB \citep{Cyburt2010}, plus additional
tabulated weak reaction rates \citet{Fuller1985, Oda1994,
  Langanke2000}. Screening is included via the prescription of
\citet{Chugunov2007}.  Thermal neutrino loss rates are from
\citet{Itoh1996}. We compute the pre-merger evolution using an
8-isotope $\alpha$-chain nuclear reaction network and switch to a
22-isotope nuclear network for the post-merger evolution.

We evolve our models from the pre-main sequence to the terminal age
main sequence of the most massive $58\,M_\odot$ star, defined as the
time when the central hydrogen abundance $X(^1\mathrm{H})\leq 10^{-4}$.
We treat convection using the Ledoux criterion, and include
thermohaline mixing (until the central temperature
$\log_{10}(T_c/\mathrm{[K]})>9.45$, \citealt{farmer:16}) and semiconvection, both with an
efficiency factor of 1. We assume $\alpha_\mathrm{MLT}=2.0$ and use
\cite{brott:11} overshooting for the convective core burning. We have
tested that varying core overshooting does not impact significantly the
post-merger evolution, however, when including shell overshooting
and/or undershooting we were unable to find solutions to the stellar
structure equations. Moreover, we employ the MLT++ artificial
enhancement of the convective flux \citep[e.g.,][]{paxton:15,
  jiang:15}. Stellar winds are included using the algorithms from
\cite{vink:01} with an efficiency factor of 1.

To compute through the very late phases, we reduce the core resolution
and increase the numerical solver tolerance when the central
temperature increases above $\log_{10}(T_c/\mathrm{[K]})>9.45$. We
define the onset of core-collapse when the iron-core infall velocity
exceeds $1000\,\mathrm{km\ s^{-1}}$ \citep[e.g.,][]{woosley:02}.

The inlists, processing scripts, and model output will be made available at~\href{https://zenodo.org/record/4062493}{https://zenodo.org/record/4062493}.

\bibliographystyle{aasjournal}
\bibliography{./merger.bib}

\end{document}